\documentclass[letter]{IEEEtran}
\usepackage{amsmath,amssymb,amsthm}
\usepackage{multirow}
\usepackage{cite}
\usepackage{epsfig}
\usepackage{color}
\usepackage{threeparttable}

\ifCLASSINFOpdf
\else
\fi
\newcommand{\defeq}{\stackrel{\text{def}}{=}}
\theoremstyle{definition}
\newtheorem{PROP}{Proposition}

\begin{document}
%
\title{Error Performance Analysis of the Symbol-Decision SC Polar Decoder}

\IEEEoverridecommandlockouts

\author{
\IEEEauthorblockN{Chenrong Xiong, Jun Lin,~\IEEEmembership{Student~member,~IEEE} and Zhiyuan Yan,~\IEEEmembership{Senior~member,~IEEE}}


}


%


\maketitle

\begin{abstract}
Polar codes are the first provably capacity-achieving forward error correction codes. To improve
decoder throughput, the symbol-decision SC algorithm
makes hard-decision for multiple bits at a time. In this paper, we prove that for
polar codes, the symbol-decision SC algorithm is better than 
the bit-decision SC algorithm in terms of the frame error rate (FER) performance because the
symbol-decision SC algorithm performs a \emph{local} maximum likelihood decoding
within a symbol. Moreover, the bigger the symbol size, the better the FER
performance. Finally, simulation results over both the additive white Gaussian
noise channel and the binary erasure channel confirm our theoretical analysis.
\end{abstract}


%
\IEEEpeerreviewmaketitle

\section{Introduction}
\label{sec:intro}

Polar codes, a groundbreaking discovery by Arikan \cite{5075875}, provably
achieve the capacity of any discrete \cite{5075875} and continuous
\cite{5351487} memoryless channels. Since their debut, a lot of efforts have
been made to improve the error performance of short polar codes. Although a
sphere decoding algorithm \cite{6283643}, stack sphere decoding algorithm
\cite{6708139} or a Viterbi algorithm \cite{ML_polar} can provide maximum
likelihood (ML) decoding of polar codes, they are considered infeasible due to their
high complexity. Compared with these ML decoding algorithms, the
successive cancellation (SC) decoding algorithm
\cite{5075875} has a lower complexity at the cost of sub-optimal
performance. Another drawback of the SC algorithm is its long decoding latency
and low decoding throughput because the SC algorithm makes hard \emph{bit} decisions
only one bit at a time. To reduce the decoding latency and improve the 
throughput from a perspective of parallel processing, a parallel SC
algorithm was proposed in \cite{ParSC} and \cite{ParSC2}. The symbol-decision SC
algorithm in \cite{SBSCL2014SiPS}, which has the same decoding schedule as the
parallel SC algorithm, makes hard \emph{symbol} decisions one at a
time. In terms of error performance, numerical simulation results in
\cite{ParSC} and \cite{ParSC2} were used to show that the parallel SC algorithm has no
performance loss compared with the SC algorithm. There is no theoretical
analysis in the literature that shows whether the parallel and symbol-decision SC algorithms
are superior or inferior to the SC algorithm \cite{5075875}, referred to as the
bit-decision SC algorithm henceforth.

In this paper, besides numerical simulations, we prove that in terms of frame
error rate (FER) performance, the symbol-decision SC algorithm is better than the
bit-decision SC algorithm. Moreover, the bigger the symbol size, the better the
FER performance. Finally, simulation results over the additive white Gaussian
noise (AWGN) channel and the binary erasure channel (BEC) confirm our
theoretical analysis.

The rest of this paper is organized as follows. In Section~\ref{sec:intro}, polar
codes are reviewed as well as the bit- and symbol-decision SC
algorithms. In Section~\ref{sec:per_analysis}, we prove that the symbol-decision
SC algorithm has a better FER performance than the bit-decision SC algorithm. In
this section, we also show how to make use of future frozen bits within a
symbol by the symbol-decision SC algorithm. Numerical simulation results are 
presented to confirm our theoretical conclusion as well. Finally, some
conclusions are provided in Section~\ref{sec:conclusion}. 

\section{Bit-Decision and Symbol-Decision SC  Algorithms for Polar Codes}
\label{sec:intro}
\subsection{Polar codes}
For simplicity, we denote $(u_a,u_{a+1},\cdots,u_{b-1},u_b)$ as $u_a^b$; if
$a>b$, $u_a^b$ is regarded as void. For any index set $\mathcal{A} \subseteq \mathcal{I}=\{1,2,\cdots, N\}$, $\mathbf{u}_{\mathcal{A}}=(u_i: 0 < i \leq N, i\in
\mathcal{A})$ is the sub-sequence of $\mathbf{u}=u_1^N$ restricted to
$\mathcal{A}$. We denote the complement of $\mathcal{A}$ in $\mathcal{I}$ as
$\mathcal{A}^c$. 

Suppose $N=2^n$, for an $(N,K)$ polar code, the data bit sequence $\mathbf{u}=u_1^N$ is
divided into two parts: a $K$-element part $\mathbf{u}_{\mathcal{A}}$ which
carries information bits, and $\mathbf{u}_{\mathcal{A}^c}$ whose elements are
predefined frozen bits. For convenience, frozen bits are set to zero. 

To generate the corresponding encoded bit sequence $\mathbf{x}=x_1^N$,
\begin{equation}
\mathbf{x} = \mathbf{u}B_NF^{\otimes n},
\label{equ:encoder}
\end{equation}
where 
 $B_N$ is the $N\times N$ bit-reversal permutation matrix, $F=\left[
\begin{smallmatrix}
1 & 0 \\
1 & 1 
\end{smallmatrix}
\right]$, and $F^{\otimes
  n}$ is the $n$-th Kronecker power of $F$ \cite{5075875}.

\subsection{Bit-Decision SC Algorithm for Polar Codes}
When $\mathbf{x}$ is transmitted, suppose the received bits are
$\mathbf{y}=y_1^N$. The bit-decision SC algorithm \cite{5075875} for an $(N,K)$ polar code estimates the
data bit sequence $\mathbf{u}$ successively: for $j=1,2,\cdots,N$,
$\hat{u}_j=0$ if $u_j$ is a frozen bits, otherwise it is estimated by
\begin{equation}
\label{eq:SC_DR}
\hat{u}_j=\underset{u_j\in \{0,1\}}{\arg\max}\Pr(\mathbf{y},\hat{u}_1^{j-1}|u_j).
\end{equation} 
Here, the bit-decision SC algorithm makes hard bit decisions one bit at a time.

\subsection{Symbol-Decision SC Algorithm for Polar Codes}

%
The $M$-bit\footnote{Although the symbol size $M$ can be any integer no more
  than $N$, we assume $M\vert N$ for simplicity.} parallel and symbol-decision SC algorithms \cite{ParSC, ParSC2, SBSCL2014SiPS} make
hard-decision for $M$ bits instead of only one bit at a time. For $0 \leq j <
\frac{N}{M}$, the $j$-th symbol is estimated successively by
\begin{equation}
\label{eq:MB_DR}
\hat{u}_{jM+1}^{jM+M} =
\underset{\substack{
\mathbf{u}_{\mathcal{AM}_j}\in\{0,1\}^{\lvert\mathcal{AM}_j\rvert} \\
\mathbf{u}_{\mathcal{AM}_j^c}\in\{0\}^{\lvert\mathcal{AM}_j^c\rvert}}}{\arg\max}{\Pr}(\mathbf{y},\hat{u}_1^{jM}|u_{jM+1}^{jM+M}),
\end{equation}
where $\mathcal{IM}_j \defeq
\{jM+1,jM+2,\cdots,jM+M\}\subseteq \mathcal{I} $, $\mathcal{AM}_j \defeq \mathcal{IM}_j \cap
\mathcal{A}$, $\mathcal{AM}_j^c \defeq \mathcal{IM}_j \cap \mathcal{A}^c$, and
$\lvert\mathcal{AM}_j\rvert$ represents the cardinality of 
$\mathcal{AM}_j$. If $M=N$, the $M$-bit symbol-decision
SC algorithm is exactly an ML sequence decoding algorithm. 

\section{Performance Analysis of the Symbol-Decision SC Algorithm}
\label{sec:per_analysis}

\subsection{FER Analysis of the Symbol-Decision SC Decoding
  Algorithm}
\begin{figure}[htbp]
\centering
\includegraphics[width=6.5cm]{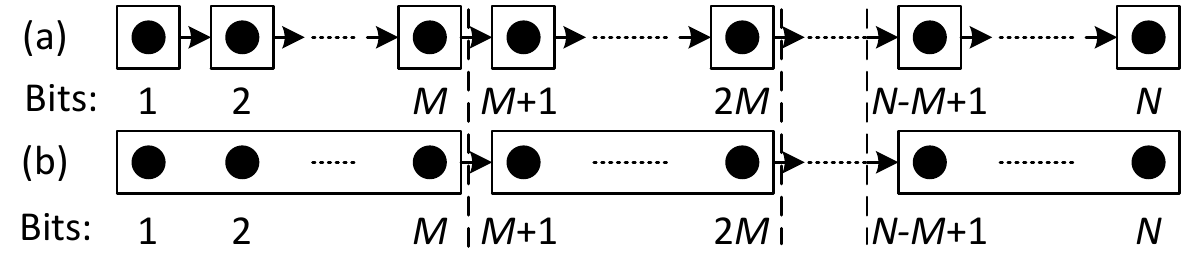}
\caption{Decoding procedures of (a) a bit-decision SC algorithm and (b) an
  $M$-bit symbol-decision SC algorithm.}
\label{fig:SCvsML}
\end{figure}

To have a fair comparison, we assume the symbol-decision decoding has the same
bit sequence $\mathbf{u}$
 as its bit-decision counterpart. Without loss of generality, we
consider two decoding scenarios shown in 
Fig.~\ref{fig:SCvsML}. In both scenarios, an $N$-bit vector is divided into 
$\frac{N}{M}$ segments. Each segment has $M$ bits. The bit-decision SC and $M$-bit ML
decoding algorithms are used to decode each segment of scenarios (a) and (b),
respectively. A box means that a decision is made. From a segment to the following segment, both scenarios use the
same schedule -- the successive schedule. Then scenarios (a) and (b) exactly
correspond to the bit-decision SC and $M$-bit symbol-decision SC algorithms,
respectively. Note that when a different bit sequence is used for both, all
conclusions still apply.

In terms of the FER performance, we have
 
\begin{PROP}
\label{prop:msbbs}
If all data sequences are independent and equally likely, for an $(N,K)$ polar
code over any given channel, the FER of the
bit-decision SC algorithm ${\Pr}_{\text{B}}(\hat{u}_1^{N} \neq
u_1^{N})$ and the FER of the $M$-bit symbol-decision SC
algorithm ${\Pr}_{\text{M}}(\hat{u}_1^{N} \neq u_1^{N})$ satisfy:

\begin{equation}
\label{eq:SLB}
{\Pr}_{\text{M}}(\hat{u}_1^{N} \neq u_1^{N}) \leq
{\Pr}_{\text{B}}(\hat{u}_1^{N} \neq u_1^{N}).
\end{equation}
\end{PROP}

\begin{IEEEproof}
Let us calculate the FERs of the two
scenarios shown in Fig.~\ref{fig:SCvsML}. Let $p_0={\Pr}_{\text{SC}}(\hat{u}_{1}^{M} \neq u_{1}^{M})$ and
$p_0^{\prime}={\Pr}_{\text{ML}}(\hat{u}_{1}^{M}\neq u_{1}^{M})$ denote the
segment error rate of $\hat{u}_{1}^{M}\neq u_{1}^{M}$ by using the SC and ML
decoding algorithms, respectively. Similarly, for $i=1,2,\cdots,\frac{N}{M}-1$, let $p_i={\Pr}_{\text{SC}}(\hat{u}_{iM+1}^{iM+M} \neq
u_{iM+1}^{iM+M}|\hat{u}_1^{iM}=u_1^{iM})$ and
$p_i^{\prime}={\Pr}_{\text{ML}}(\hat{u}_{iM+1}^{iM+M}\neq
u_{iM+1}^{iM+M}|\hat{u}_1^{iM}=u_1^{iM}) (1\leq i < \frac{N}{M})$ 
represent the probabilities of that the $i$-th segment is erroneously decoded by
the SC and ML decoding algorithms, respectively, provided that all previous
segments are correctly decoded.

Then we have  the segment error probability ${\Pr}(\hat{u}_{1}^{M}\neq
u_{1}^{M}) = \sum_{y_1^N}{\Pr}(\hat{u}_{1}^{M}\neq
u_{1}^{M}|y_1^N){\Pr}(y_1^N)$. Since ${\Pr}(y_1^N)$ is independent of the
decoding rule, to
minimize ${\Pr}(\hat{u}_{1}^{M}\neq u_{1}^{M})$, we need to minimize
${\Pr}(\hat{u}_{1}^{M}\neq u_{1}^{M}|y_1^N)$, i.e., to maximize
${\Pr}(\hat{u}_{1}^{M} = u_{1}^{M}|y_1^N)$.

Because
\begin{equation*}
{\Pr}(u_{1}^{M}|y_1^N) = \frac{{\Pr}(y_1^N|u_{1}^{M}){\Pr}(u_{1}^{M})}{{\Pr}(y_{1}^{N})},
\end{equation*}
and $u_{1}^{M}$ is a uniformly distributed random variable, the ML decoder
maximizes ${\Pr}(\hat{u}_{1}^{M} = u_{1}^{M}|y_1^N)$. Therefore, we have
\begin{equation}
\label{eq:i0}
p_0 \geq p_0^{\prime}.
\end{equation} 

For any $1 \leq i < \frac{N}{M}$, the segment error probability ${\Pr}(\hat{u}_{iM+1}^{iM+M}\neq
u_{iM+1}^{iM+M}|\hat{u}_{1}^{iM}=u_{1}^{iM}) = \sum_{y_1^N}{\Pr}(\hat{u}_{iM+1}^{iM+M}\neq
u_{iM+1}^{iM+M}|y_1^N,\hat{u}_{1}^{iM}=u_{1}^{iM}){\Pr}(y_1^N)$. Hence, to minimize ${\Pr}(\hat{u}_{iM+1}^{iM+M}\neq
u_{iM+1}^{iM+M}|\hat{u}_{1}^{iM}=u_{1}^{iM})$, ${\Pr}(\hat{u}_{iM+1}^{iM+M}=
u_{iM+1}^{iM+M}|y_1^N,\hat{u}_{1}^{iM}=u_{1}^{iM})$ need to be maximized.

Because 
\begin{equation*}
\begin{split}
{\Pr}&(u_{iM+1}^{iM+M}|y_1^N,\hat{u}_{1}^{iM}=u_{1}^{iM})=\\
&\frac{{\Pr}(y_1^N,\hat{u}_{1}^{iM}=u_{1}^{iM}|u_{iM+1}^{iM+M}){\Pr}(u_{iM+1}^{iM+M})}{{\Pr}(y_{1}^{N},\hat{u}_{1}^{iM}=u_{1}^{iM})},
\end{split}
\end{equation*}
and $u_{iM+1}^{iM+M}$ is a uniformly distributed random variable, the ML decoder
maximizes ${\Pr}(\hat{u}_{iM+1}^{iM+M} =
u_{iM+1}^{iM+M}|y_1^N,\hat{u}_{1}^{iM}=u_{1}^{iM})$. Therefore, we also have
\begin{equation}
\label{eq:SCLML}
p_i \geq p_i^{\prime}\text{ for } 1 \leq i < \frac{N}{M}.
\end{equation}

For the bit-decision SC algorithm, 
\begin{equation*}
{\Pr}_{\text{B}}(\hat{u}_1^{N} \neq u_1^{N}) =1-\prod_{i=0}^{\frac{N}{M}-1}(1-p_i).
\end{equation*}

For the $M$-bit symbol-decision SC algorithm, 
\begin{equation*}
{\Pr}_{\text{M}}(\hat{u}_1^{N} \neq u_1^{N})  =1-\prod_{i=0}^{\frac{N}{M}-1}(1-p_i^{\prime}).
\end{equation*}

According to \eqref{eq:i0} and \eqref{eq:SCLML}, we have 
\begin{equation*}
{\Pr}_{\text{M}}(\hat{u}_1^{N} \neq u_1^{N}) \leq
{\Pr}_{\text{B}}(\hat{u}_1^{N} \neq u_1^{N}).
\end{equation*}
\end{IEEEproof}

\begin{figure}[htbp]
\centering
\includegraphics[width=7.5cm]{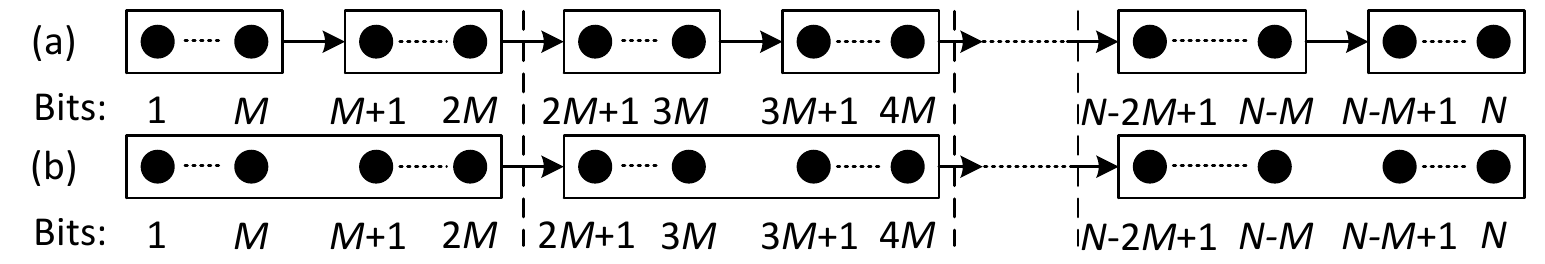}
\caption{Decoding procedures of (a) an $M$-bit symbol-decision SC algorithm and (b) a
  $2M$-bit symbol-decision SC algorithm.}
\label{fig:MLvs2ML}
\end{figure}

\begin{figure*}[htbp]
\centering
\includegraphics[width=11cm]{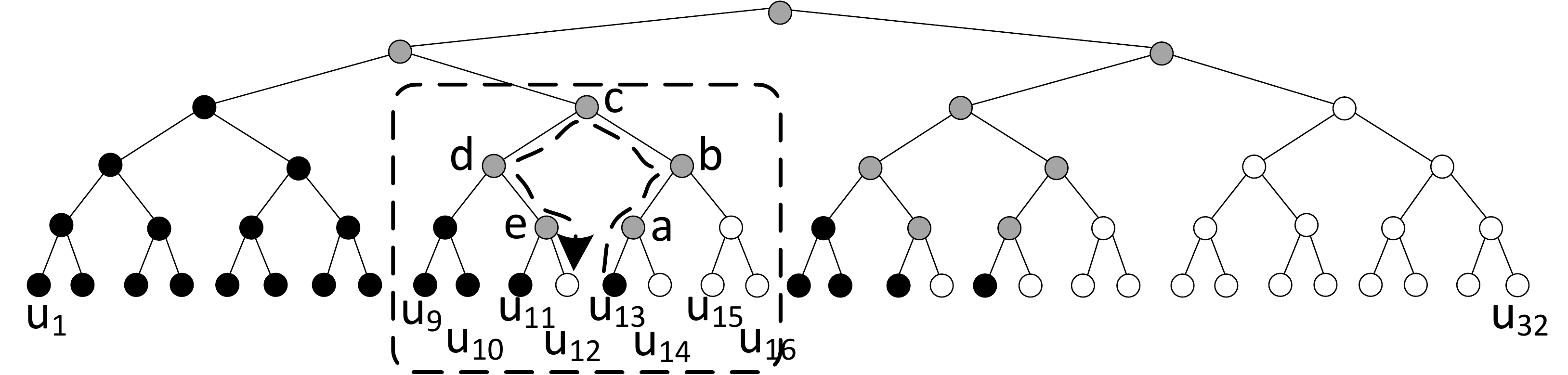}
\caption{Tree graph of a (32, 16) polar code.}
\label{fig:TG}
\end{figure*}

Furthermore, we have
\begin{PROP}
\label{prop:2MsbbMs}
If all data sequences are independent and equally likely, for an $(N,K)$ polar code over any given channel, the FER of an
$M$-bit symbol-decision SC algorithm ${\Pr}_{\text{M}}(\hat{u}_1^{N} \neq
u_1^{N})$ and the FER of a $2M$-bit symbol-decision SC
algorithm ${\Pr}_{\text{2M}}(\hat{u}_1^{N} \neq u_1^{N})$ satisfy:
\begin{equation}
\label{eq:2SLB}
{\Pr}_{\text{2M}}(\hat{u}_1^{N} \neq u_1^{N}) \leq
{\Pr}_{\text{M}}(\hat{u}_1^{N} \neq u_1^{N}).
\end{equation}
\end{PROP}

By considering the two scenarios in Fig.~\ref{fig:MLvs2ML},
Proposition~\ref{prop:2MsbbMs} can be proved in a similar way as for Proposition~\ref{prop:msbbs}.

Therefore, the symbol-decision SC algorithm is no worse than
the bit-decision SC algorithm in terms of the FER performance and bridges the
FER performance gap between the bit-decision SC algorithm and the ML decoding algorithm. 

\subsection{Message Passing Interpretation}

The SC algorithm can be considered as message passing over a tree graph
\cite{6065237}. From the perspective of message passing over a tree graph, we provide
an explanation of the advantage of the symbol-decision decoding. To this end, 
we introduce a string vector $\mathbb{S}_i=$'$\mathcal{S}_{i,1},\cdots,\mathcal{S}_{i,M}$' (for $0 \leq i <
\frac{N}{M}$) to represent a data pattern of the $i$-th $M$-bit symbol of a polar
code with length $N$. If $u_{iM+j}$ is an information bit, $S_{i,j}$ is
denoted as '$\mathcal{D}$'. Otherwise, $S_{i,j}$ as '$\mathcal{F}$'. Consider a toy example of a 4-bit symbol $u_{4i+1}^{4i+4}$. Assuming $u_{4i+1}$ and $u_{4i+3}$
are information bits, and $u_{4i+2}$ and $u_{4i+4}$ are frozen bits. Then the data pattern of
$u_{4i+1}^{4i+4}$ is '$\mathcal{DFDF}$'. Obviously, for an $M$-bit symbol, there are $2^M$ possible data patterns. We
divide them into two types. The first type is called a DP-I pattern, which has
no '$\mathcal{D}$' or has no '$\mathcal{F}$' after the first
'$\mathcal{D}$'. There are only $(M+1)$ DP-I patterns. The remaining $(2^M-M-1)$
patterns are called DP-II patterns. Henceforth, a symbol which has a DP-I
(DP-II, respectively) pattern is called a DP-I (DP-II, respectively) symbol. 

As pointed out in \cite{5075875}, the bit-decision decoding does not take advantage
of future frozen bits. That is, when decoding information bit $u_i\text{ }(i\in
\mathcal{A})$, the fact that $u_j\text{ }(j\in \mathcal{A}^c\text{ and }j>i)$ is
a frozen bit is not accounted for by the bit-decision SC algorithm. For the symbol-decision SC algorithm, the
future frozen bits in future symbols and within a DP-I symbol cannot be taken advantage of either. However, the decision
rule of the symbol-decision SC algorithm can be regarded as a \emph{local} ML decoder. As a result, some
information bits can take advantage of their future frozen bit(s) within any DP-II
symbol. 

We consider a tree graph representation, shown in Fig.~\ref{fig:TG}, of a (32,
16) polar code constructed with the method in \cite{4542778}.
Nodes on the bottom (from left to right, $u_1$ to $u_{32}$) are called leaf
nodes. Each leaf node corresponds to a data bit. There are three kinds of nodes
in the tree graph. A rate-0 node whose descendant leaf nodes are all frozen
bits is represented by a black node. A rate-1 node whose descendant leaf nodes are
all information bits is represented by a white node. The rest are rate-R nodes in gray. Some
descendant leaf nodes of a rate-R node are frozen bits, and the others are
information bits. We consider how to use the knowledge of a frozen
bit from the perspective of message passing. The knowledge of a frozen bit can be
passed through only the rate-0 nodes according to the encoding of polar codes.

Given a tree graph and $M$, data patterns are determined. For the tree
graph in Fig.~\ref{fig:TG}, all data patterns of $M=2,4$, and $8$ are listed in
Table~\ref{tab:DP}. When $M=2$ and $4$, there are no DP-II symbols.
\begin{table}[htbp]
\begin{center}
\caption{Data patterns of the (32, 16) polar code for different $M$s.}
\label{tab:DP}
\begin{tabular}{|c|c|c|}
\hline
$M$ & DP-I & DP-II \\ \hline
$2$ & $\mathcal{FF}$, $\mathcal{FD}$, $\mathcal{DD}$ & none \\ \hline
$4$ & $\mathcal{FFFF}$, $\mathcal{FFFD}$, $\mathcal{FDDD}$, $\mathcal{DDDD}$ & none
\\ \hline
$8$ & $\mathcal{FFFFFFFF}$, $\mathcal{DDDDDDDD}$ & $\mathcal{FFFDFDDD}$ \\ \hline
\end{tabular}
\end{center}
\end{table}

\begin{figure}[htbp]
\centering
\includegraphics[width=8cm]{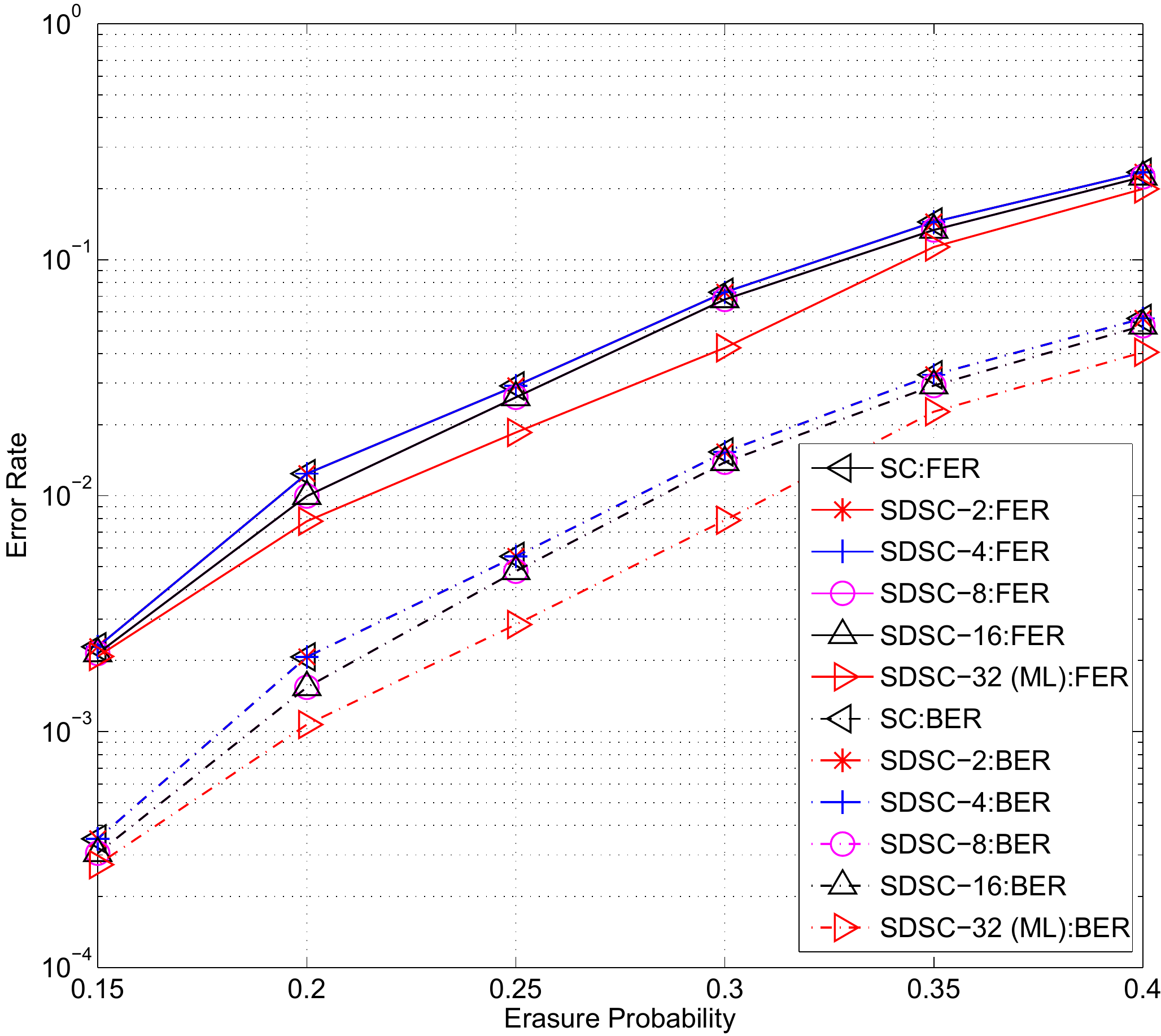}
\caption{Error rates of decoding algorithms for the (32, 16) polar code
  over the BEC.}
\label{fig:par_SC_per_32_16_BEC}
\end{figure}

Let us take the decoding of $u_{12}$
as an example. Although $u_{13}$ is a frozen bit, this knowledge needs to pass
through some intermediate nodes
a$\rightarrow$b$\rightarrow$c$\rightarrow$d$\rightarrow$e before being 
received by $u_{12}$ if it is to be taken advantage of in the decoding of
$u_{12}$. However, because there is at least one rate-R node in the message 
passing route from $u_{13}$ to $u_{12}$, the decoding of $u_{12}$ cannot take
advantage of the frozen bit $u_{13}$. However, for the 8-bit symbol-decision
SC algorithm, the DP-II symbol $u_9^{16}$ is decoded as a symbol
simultaneously. The frozen bits ($u_9^{11}$ and $u_{13}$) help to
decode the information bits ($u_{12}$ and $u_{14}^{16}$). Therefore,
unlike the bit-decision SC decoding algorithm, the 8-bit symbol-decision SC
decoding does take advantage of $u_{13}$ to decode $u_{12}$. If the 2-bit or 4-bit symbol-decision algorithm are used, no future
frozen bits can be taken advantage of in decoding any information bit because
all 2-bit or 4-bit symbols are DP-I symbols. In terms of the FER over the
BEC, SDSC-32 (ML) $<$ SDSC-16 $\approx$ SDSC-8 $<$ SDSC-4 $\approx$ SDSC-2
$\approx$ SC (shown in Fig.~\ref{fig:par_SC_per_32_16_BEC}), where SDSC-$i$
represents the $i$-bit symbol-decision SC algorithm and SDSC-32 is also an ML algorithm.

\subsection{Simulation Results}

\begin{figure}[htbp]
\centering
\includegraphics[width=8cm]{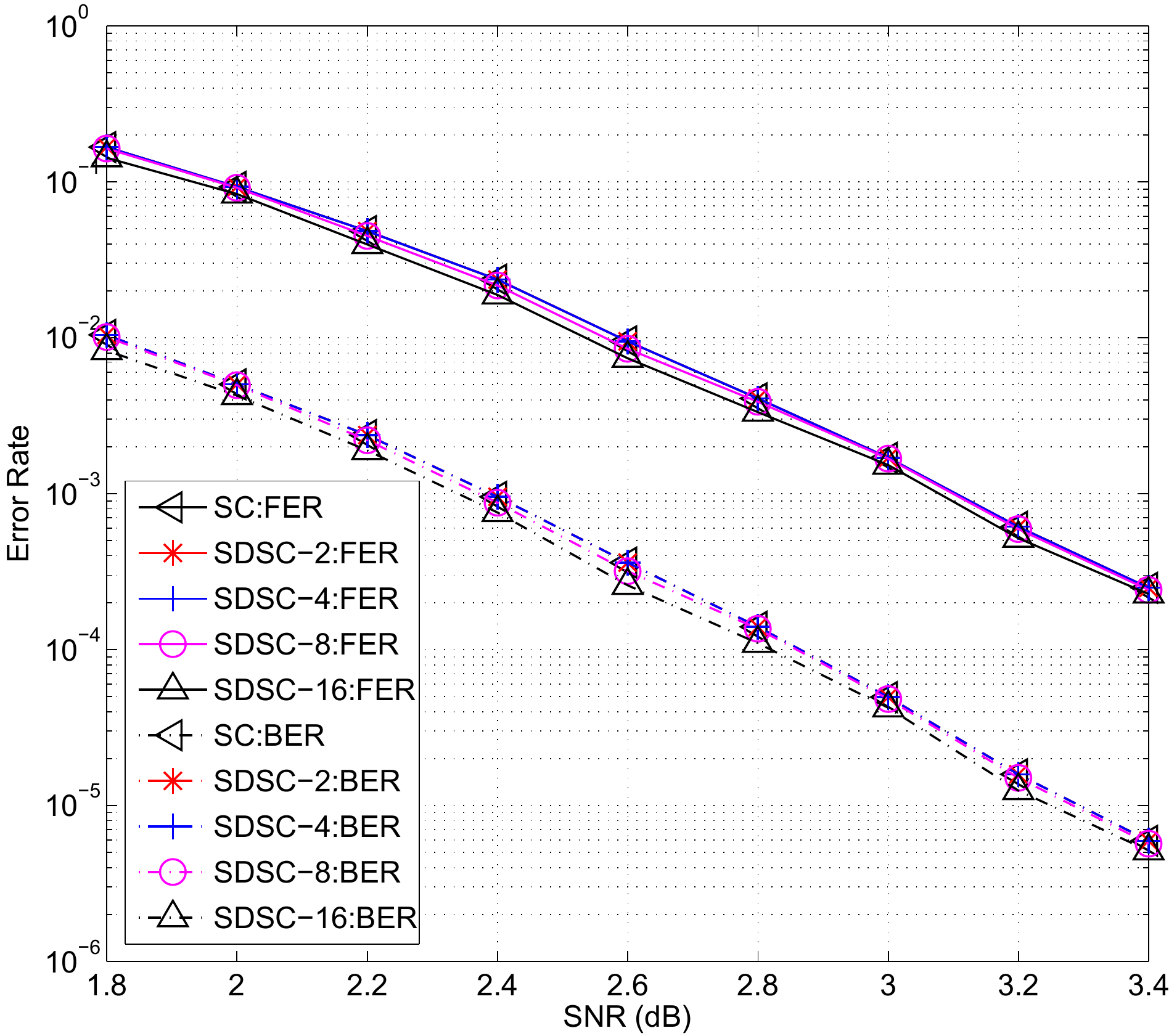}
\caption{Error rates of decoding algorithms for the (1024, 512) polar code
  over the AWGN channel.}
\label{fig:par_SC_per_1024_512_awgn}
\end{figure}

\begin{figure}[htbp]
\centering
\includegraphics[width=8cm]{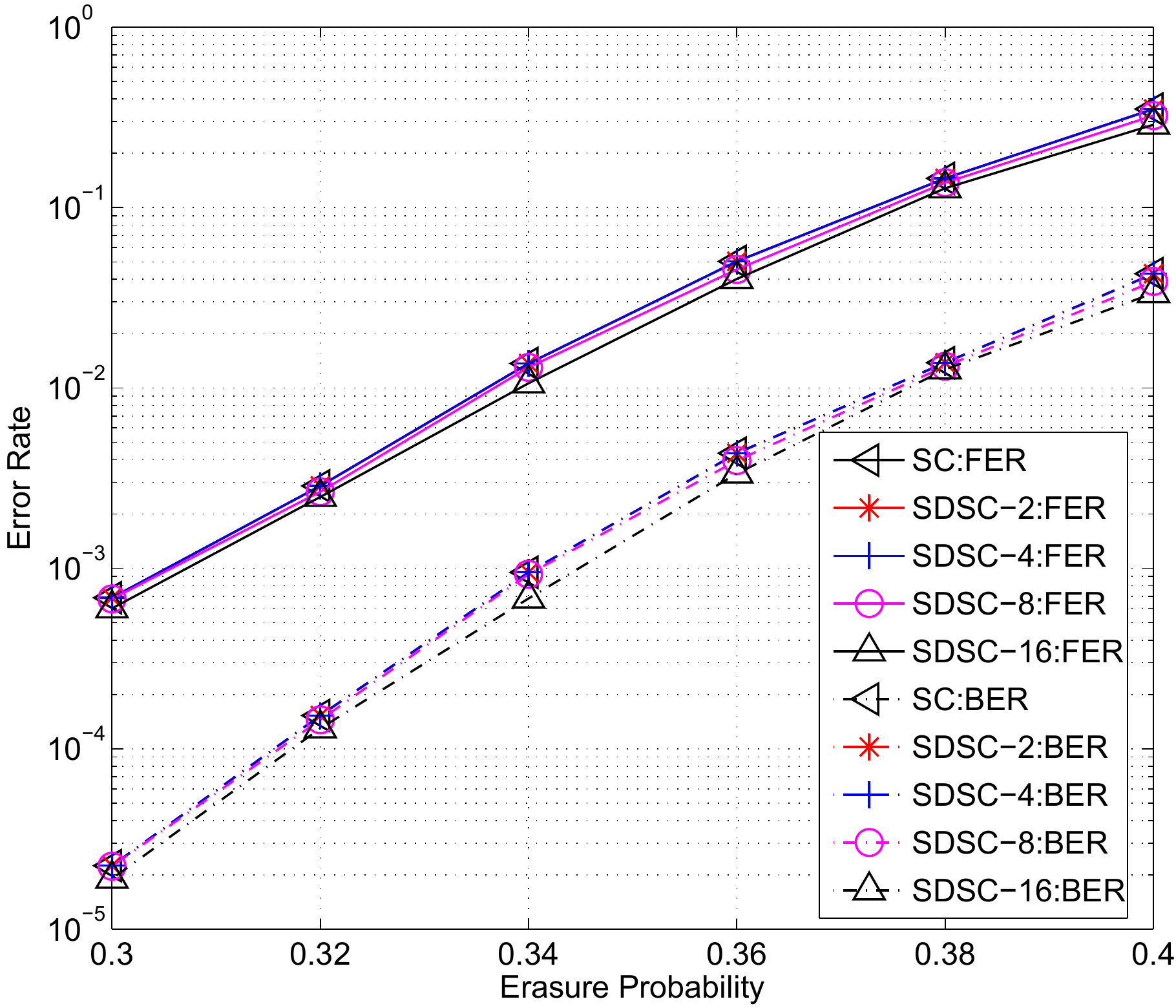}
\caption{Error rates of decoding algorithms for the (1024, 512) polar code
  over the BEC.}
\label{fig:par_SC_per_1024_512_BEC}
\end{figure}

Figs.~\ref{fig:par_SC_per_1024_512_awgn} and \ref{fig:par_SC_per_1024_512_BEC}
show bit error rates (BERs) and FERs of symbol-decision SC algorithms with different symbol sizes for a
(1024, 512) polar code constructed by the method in \cite{4542778} over the
AWGN channel and the BEC. Regarding data patterns of the (1024, 512) polar code,
all 2-bit and 4-bit data symbols are DP-I symbols. However, for the SDSC-8
algorithm, 8 of 128 data symbols are DP-II symbols. For the SDSC-16 algorithm,
12 of 64 data symbols are DP-II symbols. In terms of the FER,
SDSC-16 $<$ SDSC-8 $<$ SDSC-4 $\approx$ SDSC-2 $\approx$ SC for the (1024, 512)
polar code. The simulation results are consistent with Propositions \ref{prop:msbbs}
and \ref{prop:2MsbbMs}.

The performance gains are small in our simulation results, but these simulation results still
reveal how the symbol size affects the FER performance of the symbol-decision SC
algorithm. If a larger performance gain is expected, the symbol size should be increased
further. However, for larger symbol sizes, we do not provide the simulation results
because simulations are very time-consuming.

In terms of the BER performance, although we cannot offer a rigorous proof, we conjecture that the
symbol-decision SC algorithm is better than the bit-decision SC algorithm. The
simulation results in Figs.~\ref{fig:par_SC_per_1024_512_awgn} and
\ref{fig:par_SC_per_1024_512_BEC} are consistent with this conjecture.

\section{Conclusion}
\label{sec:conclusion}
This letter proves that the symbol-decision SC algorithm performs better than the
bit-decision SC algorithm for polar codes in terms of the FER
performance. Increasing the symbol size increases the FER performance 
gain. Therefore, the symbol-decision SC algorithm bridges the FER performance
gap between the bit-decision SC algorithm and the ML decoding algorithm for
polar codes.

%



\bibliographystyle{IEEEtran}
\bibliography{../latex/bibtex/IEEEfull,../latex/Polar}

%
%
%

\end{document}